\documentclass[aps,pre,10pt,twocolumn,superscriptaddress,longbibliography,balancelastpage,nofootinbib]{revtex4-2}

\usepackage{graphicx}
\usepackage{graphics}
\usepackage{amsmath}
\usepackage{amssymb}
\usepackage{amsfonts}
\usepackage{dsfont}
\usepackage{braket}
\usepackage{color}
\usepackage{braket,slashed}
\usepackage[mathscr]{euscript}
\definecolor{darkblue}{rgb}{0, 0, 0.8}
\usepackage[colorlinks=true, breaklinks=true, linkcolor=red, citecolor=blue, urlcolor=blue]{hyperref} 
\usepackage{hyperref}
\usepackage{subfigure}
\usepackage{xfrac}
\usepackage{bm}
\usepackage{kantlipsum}
\usepackage{enumitem}
\usepackage{tikz}
\usepackage{framed}
\usepackage{graphicx}
\usepackage{subfigure}
\usepackage{cleveref}
\usepackage{csquotes}
\usepackage{xcolor}

\allowdisplaybreaks[1]

\newcommand{\Othree}{\ensuremath{\mathrm{O}(3)}}
\newcommand{\ON}{\ensuremath{\mathrm{O}(N)}}
\newcommand{\SU}{\ensuremath{\mathrm{SU}}}
\newcommand{\SO}{\ensuremath{\mathrm{SO}}}
\newcommand{\SOthree}{\ensuremath{\mathrm{SO}(3)}}
\newcommand{\RP}{\ensuremath{\mathrm{RP}^2}}
\newcommand{\Z}{\ensuremath{\mathbb{Z}}}
\newcommand{\threej}[6]{\ensuremath{\begin{pmatrix} #1 & #2 & #3 \\ #4 & #5 & #6 \end{pmatrix}}}

\DeclareMathOperator*{\argmin}{arg\ min}
\newcommand{\cH}{\ensuremath{\mathcal{H}}}
\newcommand{\cL}{\ensuremath{\mathcal{L}}}
\newcommand{\cO}{\ensuremath{\mathcal{O}}}
\newcommand{\cP}{\ensuremath{\mathcal{P}}}
\newcommand{\cR}{\ensuremath{\mathcal{R}}}
\newcommand{\cZ}{\ensuremath{\mathcal{Z}}}
\newcommand{\eps}{\ensuremath{\varepsilon}}
\newcommand{\txd}{\ensuremath{\text{d}}}
\newcommand{\txe}{\ensuremath{\text{e}}}
\newcommand{\lmax}{\ensuremath{\ell_{\text{max}}}}

\newcommand{\diagram}[2]{\;\vcenter{\hbox{\includegraphics[scale=0.32,page=#2]{#1.pdf}}}\;}

\begin{document}

\title{Contrasting pseudo-criticality in the classical two-dimensional Heisenberg and \texorpdfstring{\RP}{} models: zero-temperature phase transition versus finite-temperature crossover}

\author{Lander Burgelman}
\email{lander.burgelman@ugent.be}
\affiliation{Department of Physics and Astronomy, Ghent University, Krijgslaan 281, 9000 Gent, Belgium}
\author{Lukas Devos}
\affiliation{Department of Physics and Astronomy, Ghent University, Krijgslaan 281, 9000 Gent, Belgium}
\author{Bram Vanhecke}
\affiliation{Department of Physics and Astronomy, Ghent University, Krijgslaan 281, 9000 Gent, Belgium}
\affiliation{Faculty of Physics, University of Vienna, Boltzmanngasse 9, 1090 Vienna, Austria}
\author{Frank Verstraete}
\affiliation{Department of Physics and Astronomy, Ghent University, Krijgslaan 281, 9000 Gent, Belgium}
\author{Laurens Vanderstraeten}
\affiliation{Department of Physics and Astronomy, Ghent University, Krijgslaan 281, 9000 Gent, Belgium}


\begin{abstract}
Tensor-network methods are used to perform a comparative study of the two-dimensional classical Heisenberg and $ \RP $ models. We demonstrate that uniform matrix product states (MPS) with explicit $\SOthree$ symmetry can probe correlation lengths up to $\cO(10^3)$ sites accurately, and we study the scaling of entanglement entropy and universal features of MPS entanglement spectra. For the Heisenberg model, we find no signs of a finite-temperature phase transition, supporting the scenario of asymptotic freedom. For the $\RP$ model we observe an abrupt onset of scaling behavior, consistent with hints of a finite-temperature phase transition reported in previous studies. A careful analysis of the softening of the correlation length divergence, the scaling of the entanglement entropy and the MPS entanglement spectra shows that our results are inconsistent with true criticality, but are rather in agreement with the scenario of a crossover to a pseudo-critical region which exhibits strong signatures of nematic quasi-long-range order at length scales below the true correlation length. Our results reveal a fundamental difference in scaling behavior between the Heisenberg and $\RP$ models: Whereas the emergence of scaling in the former shifts to zero temperature if the bond dimension is increased, it occurs at a finite bond-dimension independent crossover temperature in the latter.
\end{abstract}

\maketitle


\section{Introduction}
\label{sec:intro}

The classical $ \ON $ models are among the most studied statistical mechanics models and serve as paradigmatic examples of different types of critical behavior of spin systems at equilibrium \cite{pelissetto2002critical}. Especially the two-dimensional versions remain of particularly great interest: Whereas the two-dimensional Ising model ($ N = 1 $) serves as the prototypical system exhibiting a symmetry breaking phase transition, this symmetry breaking is forbidden for all models with $ N > 1 $ by virtue of the Mermin-Wagner theorem \cite{mermin1966absence}. This, however, does not rule out a finite-temperature phase transition, as for the XY ($ N = 2 $) model a topological Berezinsky-Kosterlitz-Thouless (BKT) transition separates the gapped phase from a low-temperature critical phase \cite{berezinsky1972destruction, kosterlitz1973ordering}. The $ N = 3 $ Heisenberg model on the other hand is thought to exhibit asymptotic freedom, making it an ideal testing ground for uncovering some of the nonperturbative aspects of quantum chromodynamics.

In this work we investigate the latter model in a comparative study with a modification thereof, the $ \RP $ model \cite{lebwohl1972nematicliquidcrystal}, which was originally introduced to capture the isotropic-nematic transition in liquid crystals. Most known results for either of these models have been obtained using high-temperature expansions, Monte Carlo simulations and field-theoretical perturbative renormalization group (RG) treatments. However, despite an extensive list of works on the topic, several fundamental questions remain largely unsettled. The purpose of this work is to shed a new light on one of these questions, namely whether or not the two-dimensional $ \RP $ model exhibits a finite-temperature phase transition to a critical low-temperature phase with quasi-long-range order (QLRO). To achieve this we make use of an entanglement-based approach which provides access to aspects of these models in an entirely different manner from the established methods mentioned above, namely the framework of tensor networks \cite{verstraete2008matrix,bridgeman2017handwaving,cirac2021matrix}. This approach gives a precise characterization of the entanglement structure of the leading eigenvectors of the transfer matrix, works directly in the thermodynamic limit and allows, in combination with appropriate entanglement-scaling techniques, for a fine-grained determination of the critical behavior in these models.

Following the recent successful tensor-network studies of the XY model \cite{yu2014tensor, vanderstraeten2019approaching, ueda2021resolving}, 
we use the formalism of boundary matrix product states (MPS) \cite{haegeman2017diagonalizing, fishman2018faster, vanderstraeten2019tangentspace} for the variational characterization of the leading eigenvalue and eigenvector of the row-to-row transfer matrix, and make use of the theory of entanglement scaling \cite{tagliacozzo2008scaling, pollmann2009theory, pirvu2012matrix, rams2018precise, vanhecke2019scaling} to characterize the scaling behavior of the Heisenberg and $\RP$ models in the low-temperature region. A similar approach was recently pursued for the case of the Heisenberg model in Ref.~\onlinecite{schmoll2021classical}.

The outline of this paper is as follows. We start with an overview of the Heisenberg and $\RP$ models in Sec.~\ref{sec:models}. We proceed in Sec.~\ref{sec:method} by writing the partition function of both models as a two-dimensional tensor network, as well as providing some details on our boundary-MPS approach for its contraction. In Sec.~\ref{sec:observables} we establish the validity of our approach by computing thermodynamic quantities in both models and comparing to known Monte-Carlo results. Section \ref{sec:correlation} details a study of the correlation length, confirming a drastic difference in its qualitative behavior for the two models. This observation is explored further in Sec.~\ref{sec:entanglement}, where it is characterized in terms of a fundamental difference in scaling behavior of the entanglement entropy. Combined with an investigation of the entanglement spectrum, these results lead to a characterization of the observed signatures of QLRO in the $\RP$ model. We conclude in Sec.~\ref{sec:discussion} by summarizing our results and providing some perspective on future related work.


\section{Models}
\label{sec:models}

The two-dimensional Heisenberg model is comprised of three-component classical spins of unit length placed on the sites of a square lattice which are subject to a nearest-neighbor interaction of the form 
\begin{equation}\label{eq:heis}
	\cH_{\text{Heis}} = -\sum_{\braket{ij}} \vec{s}_i \cdot \vec{s}_j\,.
\end{equation}
This model possesses a global $ \Othree $ symmetry, as its Hamiltonian is invariant under rotations and reflections of all spins. It is generally accepted, based on a perturbative RG description of the zero-temperature continuum limit, that this model is asymptotically free with a nonperturbatively generated mass gap \cite{polyakov1975interaction, brezin1976renormalization}. Analytic expressions for the exponential divergence of the correlation length near the zero-temperature fixed point have been put forward \cite{zinn-justin2002quantum}, and these results have been previously confirmed by Monte-Carlo studies to some degree \cite{shenker1980monte, fukugita1983continuum, wolff1990asymptotic}. Yet, it has been noted that truly good agreement with asymptotic scaling can only occur at very large scales \cite{caracciolo1995asymptotic}, making the direct verification of the presence of asymptotic freedom in numerical approaches a challenging task. Moreover, some conflicting results questioning the validity of the perturbative RG results with respect to the actual lattice model have surfaced over the past decades \cite{patrascioiu1992phase, patrascioiu1995superinstantons, patrascioiu2001quasiasymptotic, aguado2004new, kapikranian2007quasilongrange}.

The two-dimensional $ \RP $ or Lebwohl-Lasher model \cite{lebwohl1972nematicliquidcrystal} consists of a modification of the Heisenberg interaction to the form
\begin{equation} \label{eq:RP2}
	\cH_{\RP} = -\sum_{\braket{ij}} \left(\vec{s}_i \cdot \vec{s}_j\right)^2\,,
\end{equation}
which possesses an additional local reflection symmetry in addition to the global $ \Othree $ symmetry. This local symmetry effectively reduces the phase space of the model to the real projective plane, $ \RP $. As the $ \RP $ manifold has a nontrivial first homotopy group, $ \pi_1(\RP) = \Z_2 $, the model hosts stable topological defects, in contrast to the Heisenberg model. Whereas the phase diagram of the Heisenberg model is more or less agreed upon, the situation for the $ \RP $ model is much less established. One reason for this continuing debate originates from the three-dimensional case: From a perturbative point of view the Heisenberg and $ \RP $ models are equivalent and should therefore belong to the same universality class \cite{zinn-justin2002quantum}, yet Monte-Carlo studies have established that the three-dimensional $ \RP $ model exhibits a weakly first-order phase transition as opposed to a continuous transition which would be consistent with the three-dimensional Heisenberg universality class \cite{lebwohl1972nematicliquidcrystal}. Similarly, while the perturbative field-theory treatment predicts the two-dimensional $ \RP $ model to have a nonvanishing mass gap everywhere, if $ \Z_2 $ vortices correspond to a relevant perturbation in two dimensions their existence might alter the nonperturbative properties of the model. In particular, this could open up a path to an extended critical region via a topological transition, similar to what happens in the two-dimensional XY model.

Indeed, several previous results provide evidence for a transition to a critical low-temperature phase with QLRO \cite{solomon1981vortices, fukugita1982numerical, chiccoli1988monte, kunz1992topological, farinas-sanchez2003evidence, mondal2003finite, shabnam2016existence}. At the same time, numerical arguments against QLRO have also been given, either arguing that there is no finite-temperature phase transition \cite{duane1981phase, sinclair1982monte} or pointing towards a weakly first-order transition \cite{latha2018two}. More recently, additional arguments against the occurrence of QLRO have arisen from careful considerations of scaling theory \cite{paredesv2008no, farinas-sanchez2010critical, tomita2014finitesize} and novel analytical approaches \cite{delfino2020absence, diouane2021critical}. In addition to this finite-temperature debate, there has been an associated discussion on the asymptotic scaling in the zero-temperature limit. While some authors claim that the additional local symmetry is irrelevant and therefore the zero-temperature behavior in the $\RP$ model is controlled by the $\Othree$ fixed point \cite{hasenbusch1996models, niedermayer1996question, catterall1998nature}, others argue that $\Z_2$ vortices constitute a relevant perturbation leading to the existence of a distinct zero-temperature $\RP$ fixed point \cite{caracciolo1993new, caracciolo1993possible, tomita2014finitesize, bonati2020asymptotic}.

While the direct verification of the asymptotic zero-temperature behavior of the $\RP$ model falls beyond the scope of this work, we will offer a characterization of the signatures of criticality observed in numerical studies of the low-temperature region. In relation to this problem we mention here two relevant models which exhibit similar behavior. The first is the classical fully-frustrated antiferromagnetic Heisenberg model on the triangular lattice, which also hosts stable topological $\Z_2$ defects. Here too, it was established that the nontrivial topological content has a dramatic effect on the behavior of the system \cite{caffarel2001spin}. While previous works have alluded to a vortex-mediated low-temperature phase transition \cite{kawamura1984phase, kawamura2010z2vortex}, there remains uncertainty whether this constitutes a phase transition in the true sense \cite{southern1993spin, wintel1995monte}. In particular, while the correlation length is thought to be finite everywhere, it was shown to be enormous near the conjectured transition point \cite{okubo2021possibility}, making it difficult to distinguish from criticality. Secondly, there is the quantum bilinear-biquadratic spin-1 Heisenberg chain, for which there is a related discussion on the possible occurrence of nematic QLRO in the vicinity of the $\SU(3)$ point of the phase diagram. While the general consensus leans towards the absence of a nematic phase, possible explanations for the observed signatures of criticality will prove highly relevant to the interpretation of our results \cite{lauchli2006spin, hu2014berryphaseinduced, dai2022absence}.

Indeed, it has been suggested that the contradicting results obtained for the $\RP$ model might be explained by the absence of true QLRO, but rather by a sharp finite-temperature crossover to a pseudo-critical region with an associated drastic change in vortex density \cite{paredesv2008no, farinas-sanchez2010critical, shabnam2016existence, latha2018two}, similar to what was suggested in the triangular-lattice antiferromagnet. A possible mechanism explaining the onset of this pseudo-critical behavior at a seemingly size-independent crossover temperature may be found in Ref.~\onlinecite{catterall1998nature}. Here, the authors argue that the observed pseudo-scaling is due to the proximity of the line of $\RP$ models to a novel RG trajectory and its associated fixed point which lie just outside of the model parameter space. Such a scenario would indeed give rise to the existence of a size-independent crossover temperature below which scaling behavior originates. In a similar vein, proximity to a true critical phase has been recently suggested as the cause for the observed pseudo-criticality in the bilinear-biquadratic chain \cite{dai2022absence}. Our results turn out to be consistent with such a scenario. We also note the possibility that such a fixed point outside the model parameter space may be situated at a complex coupling, as proposed in Ref.~\onlinecite{gorbenko2018walking}.


\section{Methodology}
\label{sec:method}

We start by providing a broad overview of our framework in this section; readers familiar with symmetric boundary-MPS methods may skip ahead to further sections. All technical details can be found in App.~\ref{sec:framework}.

In order to study the statistical-mechanics models in Eqs. \eqref{eq:heis} and \eqref{eq:RP2}, we must first write their partition function as the contraction of a tensor network directly in the thermodynamic limit. The relevant partition function at inverse temperature $ \beta = 1/T $ reads
\begin{equation}\label{eq:Z}
	\cZ = \left (\prod_{i} \int \frac{\txd\Omega_i}{4\pi}\right ) \left (\prod_{\braket{ij}} \txe^{\beta\left(\vec{s}_i \cdot \vec{s}_j\right)^p}\right )\,,
\end{equation}
where $ p = 1 $ or $ p = 2 $ for the Heisenberg and $ \RP $ models respectively, and $ \txd \Omega_i = \sin \theta_i \txd \theta_i \txd \phi_i $ represents the integration measure over the spin configurations at site $ i $. In order to arrive at a network contracted over discrete indices, we apply a duality transformation which maps the continuous angle variables appearing in Eq.~\eqref{eq:Z} to the irreducible representation of $ \Othree $ \cite{savit1980duality, liu2013exact}. This is achieved by performing a character expansion of the Boltzmann weights in terms of spherical harmonics
\begin{equation}\label{eq:expansion}
	\txe^{\beta\left(\vec{s}_i \cdot \vec{s}_j\right)^p} = \sum_{\ell} f_{\ell}(\beta) \sum_{m = -\ell}^{\ell} \bar{Y}_{\ell m}(\theta_i, \phi_i) Y_{\ell m}(\theta_j, \phi_j) \,.
\end{equation}
The expansion coefficients $ f_\ell(\beta) $ are defined in terms of Legendre polynomials $ P_\ell(x) $,
\begin{equation}\label{eq:fl}
	f_\ell(\beta) = 2\pi \int_{-1}^{1} \txd x \, P_\ell(x) \txe^{\beta x^p} \,,
\end{equation}
and decay rapidly with increasing angular momentum $\ell$. After performing this expansion the angle variables can be integrated out at each site individually, resulting in an expression for the partition function of the form
\begin{equation}\label{eq:Znetwork}
	\cZ = \quad \diagram{p2}{1} \,.
\end{equation}
The fundamental object in this expression is the four-leg tensor
\begin{equation}\label{eq:O}
	\diagram{p2}{2}\,,
\end{equation}
where each leg is labeled by an angular momentum and a magnetic quantum number which are contracted over to obtain the total partition function, and the arrows indicate the direction of these charges. For an explicit expression, see App.~\ref{sec:framework}.

In order to contract the partition function we consider the corresponding row-to-row transfer matrix
\begin{equation}\label{eq:T}
	T(\beta) = \quad \diagram{p2}{3}\,,
\end{equation}
which can be viewed as an operator acting on an infinite one-dimensional chain. The value of the partition function is therefore entirely determined by the leading eigenvalue $ \Lambda(\beta) $ of this operator, which scales as the number of sites per row $ \Lambda(\beta) = \lambda(\beta)^{N_x} $. A well established method for determining the leading eigenvalue of a one-dimensional transfer matrix is to approximate the corresponding eigenvector, or fixed point, as an MPS \cite{haegeman2017diagonalizing}. As we are dealing with a translation-invariant operator, we make use of a uniform MPS characterized by a single tensor $ A $ to parametrize the fixed point directly in the thermodynamic limit,
\begin{equation} \label{eq:MPS}
	\ket{\Psi(A)} = \quad \diagram{q1}{4} \,.
\end{equation}
The dimension $ D $ of the virtual legs of the tensor $ A $ is a control parameter, called the bond dimension, that affects how well this variational ansatz can capture the true physical properties of the system. Note that we have added arrows on the virtual legs as well, in anticipation of the underlying symmetry structure of the local MPS tensor. The fixed-point MPS must then satisfy
\begin{multline}\label{eq:eig}
	\diagram{p2}{5} \approx \\ \Lambda(\beta) \diagram{p2}{4} \,.
\end{multline}
For the problem at hand, the transfer matrix is Hermitian by construction. This allows us to reformulate the task of finding the optimal MPS tensor $ A $ as a variational optimization problem for the corresponding free energy density. This optimization problem can be solved efficiently using the variational uniform MPS (VUMPS) algorithm \cite{zauner-stauber2018variational, fishman2018faster, vanderstraeten2019tangentspace}.

In the tensor-network representation, the global $ \Othree $ symmetry at the level of the Hamiltonians is translated into a symmetry of the local tensor $ O $,
\begin{equation}\label{eq:UO}
	\diagram{p2}{6} = \diagram{p2}{7}\,,
\end{equation}
where $ U_g $ is the representation of an arbitrary rotation or reflection $ g \in \Othree $. The Mermin-Wagner theorem now precludes the breaking of a global continuous symmetry at finite temperature in both models, implying that the fixed point MPS of the transfer matrix must be invariant under any global rotation $ g \in \SOthree $. By virtue of the fundamental theorem of MPS \cite{perez-garcia2007matrix, cirac2021matrix}, this in turn leads to a symmetry constraint on the local MPS tensor of the form
\begin{equation}\label{eq:UAv}
\diagram{p2}{8} = \diagram{p2}{9}\,,
\end{equation}
where $ v_g $ is a (possibly projective) representation of $ \SO(3) $. Making use of the fact that the faithful representations of $\SO(3)$ correspond to the integer representations of $\SU(2)$, while its projective representations correspond to the half-integer representations of $\SU(2)$, this means that all tensors under consideration are fully $ \SU(2) $ invariant. As such, they possess a block structure where each block is labeled by the irreducible representations of $ \SU(2) $ on each leg of the corresponding tensor. By exploiting this inherent block structure and automatically taking into account the corresponding symmetry constraints the efficiency of all tensor manipulations is greatly improved, allowing access to effective bond dimensions that are out of reach for conventional dense tensor-network approaches \cite{mcculloch2002nonabelian, singh2010tensor, weichselbaum2012nonabelian}.

In order to put this framework into practice for the models at hand, one must first introduce an approximation in the representation of the partition function. Indeed, the tensor Eq.~\eqref{eq:O} has nontrivial entries for any value of the angular momenta labeling its legs, such that each leg in principle has an infinite dimension. For the purpose of numerical simulations, all indices must therefore be truncated at a certain cutoff value $ \lmax $ for the angular momentum on all legs. Due to the specific structure of the tensor $ O $ (cf. App.~\ref{sec:framework}) such an approximation does not result in a significant loss of accuracy. In particular, all results in the main text below were obtained using $ \lmax = 5 $ for the Heisenberg model and $ \lmax = 6 $ for the $ \RP $ model. In App.~\ref{sec:cutoff} we provide evidence that this cutoff is indeed sufficient to accurately capture the behavior of the true untruncated model in the considered temperature range, as well as show that a more heavily truncated model gives rise to the same universal behavior.

In our numerical analyses, we optimize several fixed-point MPSs of the transfer matrix throughout a temperature range which is compatible with our angular momentum cutoff, where for each temperature we use MPSs with a variety of bond dimensions. This is achieved by dynamically distributing the bond dimensions over charge sectors at the virtual level up to a given truncation error of the corresponding MPS. Systematically lowering this truncation error then gives rise to an increasing bond dimension, where the minimal truncation error we employ corresponds to an MPS with a maximal effective bond dimension of $ D \approx 1000 $. In order to distribute the bond dimension over the virtual level one must first decide which spin charges will be used on the virtual indices. While in principle both integer and half-integer spin charges may occur on the virtual MPS legs, we observe that using only integer spins yields superior results over the half-integer case. Motivation for this choice is provided in App.~\ref{sec:spt}. We note here that the occurrence of half-integer spin charges on the virtual level would indicate the existence of a symmetry-protected topological phase in the system, but this is therefore ruled out by our results.

Due to the explicit use of symmetries we were able to obtain all results shown in this work using a regular desktop computer with 32 CPU cores and a few 100 GB of RAM, requiring roughly a month's worth of computation time in total.


\section{Thermodynamic quantities}
\label{sec:observables}

\begin{figure}
\includegraphics[width=\columnwidth]{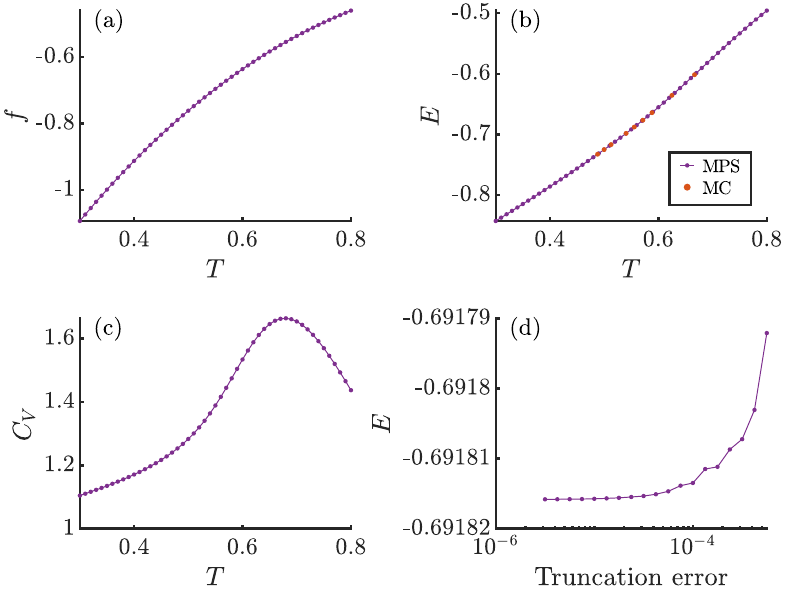}
\caption{The (a) free energy density, (b) energy per link and (c) specific heat as a function of temperature in the Heisenberg model. A comparison to the Monte-Carlo results of \cite{apostolakis1991investigation} was added in (b). (d) Convergence of the energy per link with MPS truncation error at $T = 0.55$.}
\label{fig:heisenergies_N5}
\end{figure}

\begin{figure}
\includegraphics[width=\columnwidth]{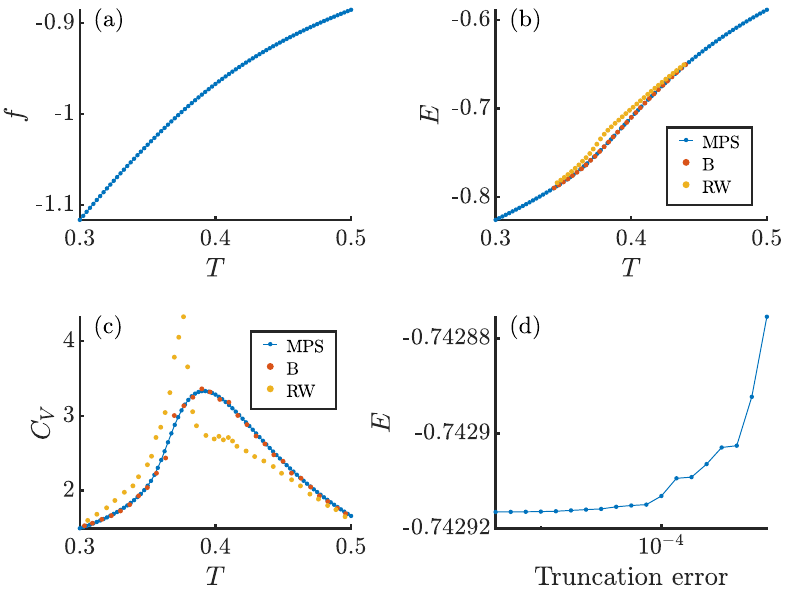}
\caption{The (a) free energy density, (b) energy per link and (c) specific heat as a function of temperature in the $\RP$ model. A comparison with the results of \cite{latha2018two} was added in (b) and (c), depicting the data obtained using both Boltzmann sampling (B) and equilibrium ensembles constructed from the density of states (RW). (d) Convergence of the energy per link with MPS truncation error at $T = 0.35$.}
\label{fig:RP2energies_N3}
\end{figure}

As a first baseline, we establish the validity of our approach by computing some thermodynamic quantities and observables for both the Heisenberg and $ \RP $ models. The way in which these quantities are obtained is detailed in App.~\ref{sec:framework}. In particular, we consider here the free energy density, the energy per link and the specific heat as a function of temperature, where the latter is computed as the temperature derivative of the energy per site. Our results are depicted in Fig.~\ref{fig:heisenergies_N5} and Fig.~\ref{fig:RP2energies_N3} for the Heisenberg and $ \RP $ models respectively.

For the Heisenberg model, we observe good agreement of the energy per link obtained with MPS compared to the Monte-Carlo results of Ref.~\onlinecite{apostolakis1991investigation}, as shown in Fig.~\ref{fig:heisenergies_N5}(b). In addition, Fig.~\ref{fig:heisenergies_N5}(d) shows excellent convergence of the energy per link with MPS truncation error, confirming the ability of uniform MPS to accurately probe local observables in the model. For the $\RP$ model we observe a similar agreement with previous works and adequate convergence of local observables. In particular, we compare the energy per link and the specific heat with the, to our knowledge, most recent Monte-Carlo results of Ref.~\onlinecite{latha2018two}. As can be seen from Figs.~\ref{fig:RP2energies_N3}(b,c) we achieve an excellent agreement of our MPS results with Monte-Carlo results obtained using conventional Boltzmann sampling, while we observe a strong deviation from the results obtained through a different sampling procedure used in \cite{latha2018two} based on constructing equilibrium ensembles from the density of states. Finally, we note that the occurrence of a distinct but rounded peak in the specific heat of the $\RP$ model has been previously attributed to the development of nematic order at short scales associated to the binding-unbinding of $\Z_2$ vortices \cite{kawamura2010z2vortex}.


\section{Correlation length}
\label{sec:correlation}

We proceed our analysis with an investigation of the correlation length, an essential quantity in diagnosing possible finite-temperature phase transitions. While MPS are generally known to faithfully capture local quantities as long as the corresponding bond dimension is increased sufficiently, recovering asymptotic properties such as the correlation length is far less straightforward. By introducing an extrapolation scheme in terms of a refinement parameter which quantifies the deviation of the inherently discrete MPS transfer matrix spectrum from a continuous one, this issue can be overcome, giving robust access to asymptotic quantities and critical properties \cite{rams2018precise, vanhecke2019scaling}. In order to extract the exact correlation length from our finite bond dimension MPS results we adopt such an extrapolation procedure introduced in Ref.~\onlinecite{rams2018precise}.

\begin{figure}
	\includegraphics[width=\columnwidth]{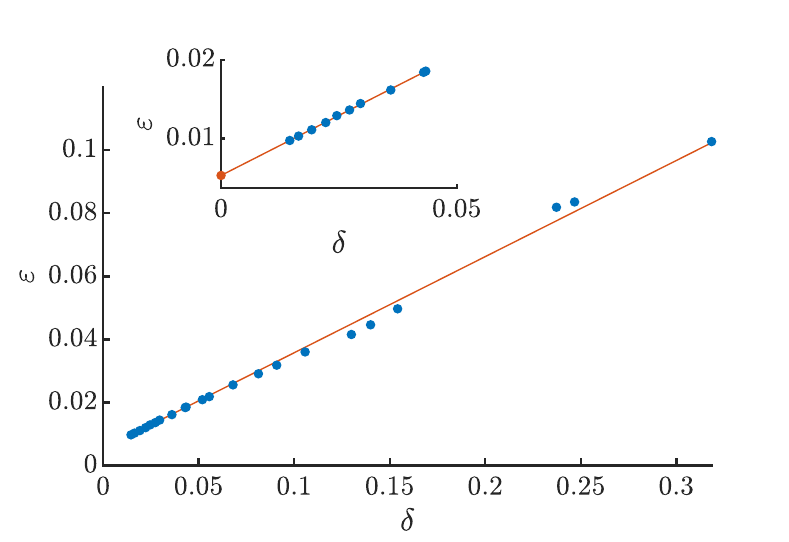}
	\caption{Illustration of the correlation length extrapolation Eq.~\eqref{eq:marek} for the $\RP$ model at $T = 0.36$. While the lowest bond dimensions clearly fall outside of the proper scaling regime, the inset shows that restricting to the 10 largest bond dimensions yields a particularly clean result of $\xi = 188 \pm 1$.}
	\label{fig:xiExtrapol}
\end{figure}

The correlation length of an MPS of bond dimension $ D $ is given by
\begin{equation}\label{eq:xi}
	\xi_D = 1/\eps, \quad \text{where} \quad \eps = -\log|\lambda_1|
\end{equation}
represents the magnitude of the second largest eigenvalue of the MPS transfer matrix, and we assume a normalized MPS with $ |\lambda_0| = 1 $. Ref.~\onlinecite{rams2018precise} details that this latter quantity scales with the gap between the second and third largest transfer matrix eigenvalues $ \delta = \log(|\lambda_1|/|\lambda_2|) $ as
\begin{equation}\label{eq:marek}
	\eps = a \, \delta^b + \eps_\infty\,,
\end{equation}
where $ \xi = 1/\eps_\infty $ is the extrapolated correlation length. For our purposes, a simple linear relation ($b = 1$) proved adequate in all applications. By restricting the eigenvalues $ \lambda_1 $ and $ \lambda_2 $ to a specific $ \SOthree $ charge sector within the symmetric tensor framework, we can directly probe the correlation length in that sector. For the Heisenberg and $ \RP $ models we will always consider correlation lengths in the $ \ell = 1 $ and $ \ell = 2 $ sectors respectively, which are the largest correlation lengths and correspond to the relevant spin-spin correlation functions in the respective models. The extrapolation procedure is illustrated for the $ \RP $ model at $ T = 0.36 $ in Fig.~\ref{fig:xiExtrapol}.

\begin{figure}
	\subfigure{\includegraphics[width=\columnwidth]{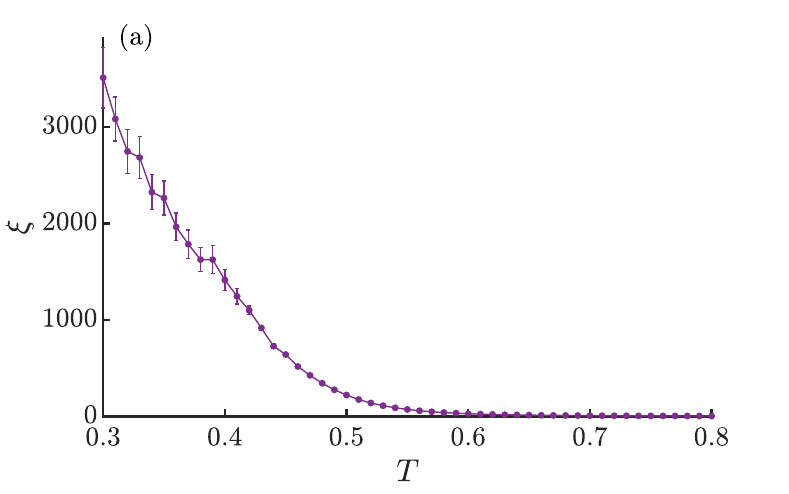}}
	\subfigure{\includegraphics[width=\columnwidth]{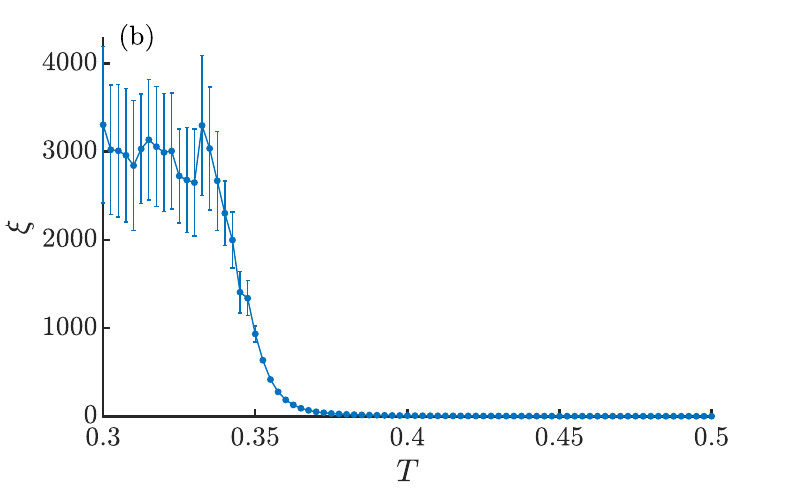}}
	\caption{Extrapolated correlation length as a function of temperature for the (a) Heisenberg and (b) $\RP$ models.}
	\label{fig:xi}
\end{figure}

Using this procedure, we extrapolate the correlation length in a temperature range compatible with our angular momentum cutoff for both the Heisenberg and $ \RP $ models. The results are depicted in Fig.~\ref{fig:xi}. We immediately note that for correlation lengths that exceed $\xi \approx 10^3$ sites, the corresponding extrapolations are quite unreliable. This is caused by the fact that the corresponding MPS fixed points are highly entangled, which results in a significant increase in the bond dimension required to access similar values of the refinement parameter $\delta$ as compared to less-entangled fixed points. As such, we are unable to accurately probe the proper scaling regime using currently accessible bond dimensions for these values. For values up to $\xi \approx 10^3$, however, extrapolation results are unchanged by an increase or decrease in maximal bond dimension used, indicating that we are able to access the proper scaling regime. Thus, we may regard values $\xi \leqslant 10^3$ as being quasi-exact, while values $\xi > 10^3$ were added to indicate changes with temperature in a qualitative manner. From Fig.~\ref{fig:xi} a stark contrast in the qualitative behavior of the correlation length with temperature in both models is immediately apparent. While for the Heisenberg model we observe a steady increase of the correlation length when lowering the temperature, the $ \RP $ model exhibits a very abrupt increase at finite temperature.

Our results for the Heisenberg model show no sign of a possible finite-temperature continuous transition in the temperature range accessible to us, in accordance with the most broadly accepted scenario that the model is gapped everywhere \cite{polyakov1975interaction, brezin1976renormalization}. From the theoretical side, an analytic prediction for the asymptotic scaling of the correlation length at finite temperature has been put forward, which can be compared to numerically obtained values to directly verify asymptotic freedom in the lattice model \cite{kim1994asymptotic}. Although the values extracted from our MPS results allow for a somewhat agreeable fit to a general exponential divergence of the correlation length at $T = 0$, they systematically fall below the theoretically predicted asymptotic scaling. This is consistent with previous results, where correlation lengths were also observed to be systematically smaller than the predicted values and it was concluded that true agreement with asymptotic scaling is only possible at very large correlation lengths \cite{kim1994asymptotic,caracciolo1995asymptotic,shin1999correction}. To quantify this statement we show in Table~\ref{tab:xi} a comparison of our MPS results, pushed to an effective bond dimension $ D \approx 3000 $ for the selected temperatures to maximize accuracy, to the Monte-Carlo results of Refs. \onlinecite{balog1999comparison} and \onlinecite{caracciolo1995asymptotic}. Our results agree perfectly with the Monte-Carlo ones for higher temperatures, whereas we seem to obtain slightly smaller correlation lengths at lower temperatures. Note that we have limited this comparison to values $\beta \leq 2.2$, as our extrapolations begin to exhibit a nonnegligible bond-dimension dependence at even lower temperatures, as discussed in the previous paragraph. Since asymptotic scaling in the temperature range investigated here would imply larger correlation lengths than those found in the Monte Carlo studies we compare to, it is clear from Table~\ref{tab:xi} that we can provide no additional evidence for the scenario of asymptotic freedom over previous works. However, as it has been established that true agreement with asymptotic scaling is only expected to occur at very low temperatures with extremely large correlation lengths which are currently inaccessible using our methods, we must conclude that our results also do not contradict this scenario. We note that our conclusions here are largely consistent with those of Ref.~\onlinecite{schmoll2021classical}, where the results of an analysis based on similar tensor-network methods leaned more towards the scenario of asymptotic freedom than that of a finite-temperature transition.

\begin{table}
    \begin{center}
        \begin{tabular}{| c || c | c |}
            \hline
            $\beta$ & $\xi_{\text{MPS}}$ & $\xi_{\text{MC}}$ \\
            \hline
            \hline
            1.5 & 11.06(0.01) & 11.04(0.01) \cite{balog1999comparison} \\
            \hline
            1.6 & 19.00(0.01) & 19.02(0.04) \cite{balog1999comparison} \\
            \hline
            1.7 & 34.51(0.01) & 34.50(0.02) \cite{balog1999comparison} \\
            \hline
            1.8 & 64.67(0.02) & 64.79(0.03) \cite{balog1999comparison} \\
            \hline
            1.9 & 121.7(0.1) & 122.3(0.1) \cite{balog1999comparison} \\
            \hline
            2.0 & 227.6(0.3) & 230.3(0.9) \cite{balog1999comparison} \\
            \hline
            2.1 & 415.1(1.8) & 422.7(2.0) \cite{caracciolo1995asymptotic} \\
            \hline
            2.2 & 727.9(4.5) & 780.0(4.8) \cite{caracciolo1995asymptotic} \\
            \hline
        \end{tabular}
    \end{center}
    \caption{Comparison of extrapolated correlation lengths for the Heisenberg model at large $\beta = 1 / T$. MPS results using effective bond dimensions up to $ D \approx 3000$ are compared to the Monte-Carlo results of Refs.~\onlinecite{balog1999comparison} and \onlinecite{caracciolo1995asymptotic}.}
    \label{tab:xi}
\end{table}

\begin{figure}
	\includegraphics[width=\columnwidth]{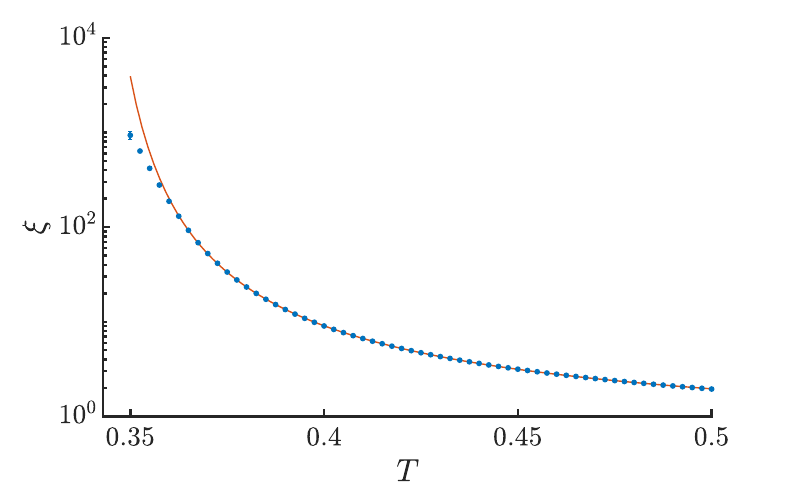}
	\caption{BKT-type fit \eqref{eq:BKTfit} to the correlation length divergence in the $\RP$ model, giving an estimate $T_c = 0.339 \pm 0.001$.}
	\label{fig:RP2kt}
\end{figure}

For the $\RP$ model the situation is drastically different. First, we observe that the correlation lengths at higher temperatures are several orders of magnitude too small to be consistent with asymptotic scaling governed by the $\Othree$ fixed point, in line with previous observations \cite{caracciolo1993possible}. In fact, at first sight our results seem to point towards a sharp divergence of the correlation length at a finite temperature. However, no single extrapolation could ever conclusively distinguish whether the correlation length is truly infinite, or rather finite but extremely large. This issue is aggravated by the fact that we are only able to reliably extrapolate correlation lengths up to $\xi \approx 10^3$. As such, it is instructive to investigate precisely how the correlation length would diverge. Many previous studies of the $ \RP $ model support the scenario of a topological phase transition driven by binding-unbinding of $\Z_2$ vortices, leading to a diverging correlation length \cite{fukugita1982numerical, kunz1992topological, farinas-sanchez2003evidence, mondal2003finite, shabnam2016existence}. However, the precise nature of the corresponding divergence in such a transition remains a matter of discussion. While our results do not allow for a satisfactory power-law fit, thereby ruling out the scenario of a second order phase transition, they are compatible with a divergence
\begin{equation}\label{eq:BKT}
	\xi \propto \exp \left(\frac{b}{\sqrt{T-T_c}}\right), \quad T \to T_c^+\,,
\end{equation}
corresponding to a possible BKT transition. Specifically, we fit a curve of the form
\begin{equation}\label{eq:BKTfit}
	\log \xi = \frac{b}{\sqrt{T-T_c}} + c + d\sqrt{T-T_c}
\end{equation}
to the extrapolated correlation lengths, where the extra terms are added to account for deviations away from the critical point. The result is shown in Fig.~\ref{fig:RP2kt}, yielding an estimated value $ T_c = 0.339 \pm 0.001 $ which is in reasonable agreement with previous estimates \cite{kunz1992topological, farinas-sanchez2003evidence, latha2018two, tomita2014finitesize, shabnam2016existence}. However, it is immediately apparent that the agreement with the form \eqref{eq:BKTfit} breaks down as the correlation length exceeds $ \xi \approx 10^2 $, above which the exponential divergence softens. For the paradigmatic XY model, a similar analysis yields an impeccable agreement with the BKT scaling form up to at least $ \xi \approx 10^3 $ \cite{vanderstraeten2019approaching}. As our extrapolations are certainly more than reliable up to these values we conclude that, in spite of the initial excellent agreement, the deviation of the system from its initial approach to a divergence provides significant evidence against a continuous transition at finite temperature. As a consequence, our correlation-length results are not consistent with the existence of a low-temperature phase with true QLRO. One could argue that the deviation we observe could be attributed to approximations, either in the angular momentum cutoff or in the truncation error of the fixed point MPS. In App.~\ref{sec:cutoff} we provide evidence that this is in fact not the case. We note that a similar deviation from an initial approach towards divergence has been observed in recent studies of the $\RP$ model \cite{tomita2014finitesize, latha2018two} as well as the triangular-lattice antiferromagnet \cite{wintel1995monte, kawamura2010z2vortex, okubo2021possibility}, where it has led to a similar conclusion.

We conclude this section by noting that even though a detailed analysis rules against a true finite-temperature divergence, the fact remains that a sudden and dramatic increase in correlation length occurs at a well-defined temperature whose value is independent of the bond dimensions used. This implies that it is a fundamental property of the two-dimensional $\RP$ model which merits an explanation. One possible explanation is to interpret this phenomenon as a sharp crossover to a pseudo-critical region due to the vicinity of a true fixed point just outside the model parameter space \cite{catterall1998nature}, as discussed at the end of Sec.~\ref{sec:models}.


\section{Entanglement properties and scaling}
\label{sec:entanglement}

Next we turn to the entanglement properties of the transfer-matrix fixed points in a more direct sense, by considering the scaling of their entanglement entropy and the nature of their entanglement spectra. This approach will further establish the distinction between the Heisenberg and $\RP$ models. In addition, it will allow us to characterize the signatures of criticality observed in the low-temperature region of the $\RP$ model in a way that is consistent with the results of the previous section.

From the study of one-dimensional quantum spin chains we know that for gapped systems the bipartite entanglement entropy of the ground state is finite. As such, the entanglement entropy of a corresponding MPS approximation will saturate as the bond dimension is increased sufficiently. In contrast, for a critical system described by a conformal field theory (CFT), the entanglement entropy of an MPS approximation to the ground state will scale as a function of its correlation length as \cite{pollmann2009theory}
\begin{equation}\label{eq:cardy}
	S_D = \frac{c}{6}\log(\xi_D) + \text{const.}\,,
\end{equation}
where $c$ is the central charge of the corresponding CFT. In more recent studies of the quantum spin-1 bilinear-biquadratic Heisenberg chain \cite{hu2014berryphaseinduced, dai2022absence} it was observed that such a scaling may also occur in a gapped system at scales below the true correlation length. That is, a system may exhibit a pseudo-critical region in which it appears critical at \enquote{small} length scales, with an entanglement-entropy scaling \eqref{eq:cardy} governed by an \emph{effective} central charge $c$. In this context the term \enquote{pseudo} denotes an effect which occurs at finite bond dimensions, but which would disappear if the bond dimension is increased until the MPS correlation length approaches the true correlation length sufficiently. Here we apply this characterization of scaling behavior in terms of an effective central charge to the Heisenberg and $\RP$ models. A subsequent analysis of the corresponding results then allows us to distinguish whether this scaling corresponds to true criticality, or rather indicates a pseudo-critical region.

\begin{figure}
	\subfigure{\includegraphics[width=\columnwidth]{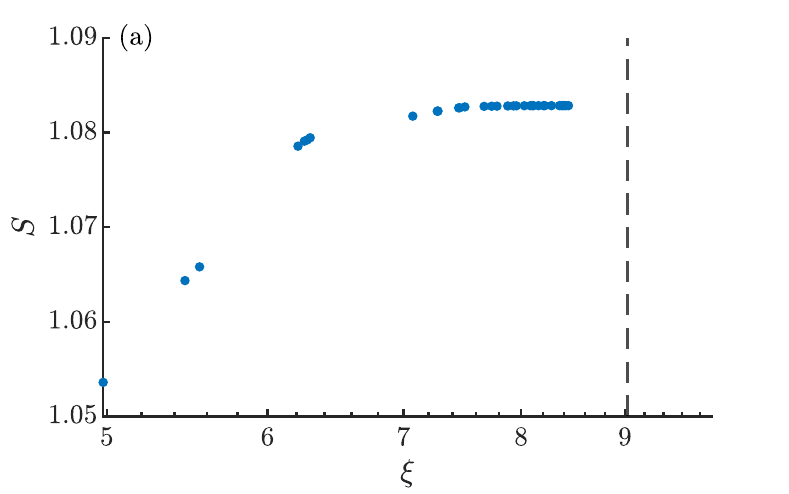}}
	\subfigure{\includegraphics[width=\columnwidth]{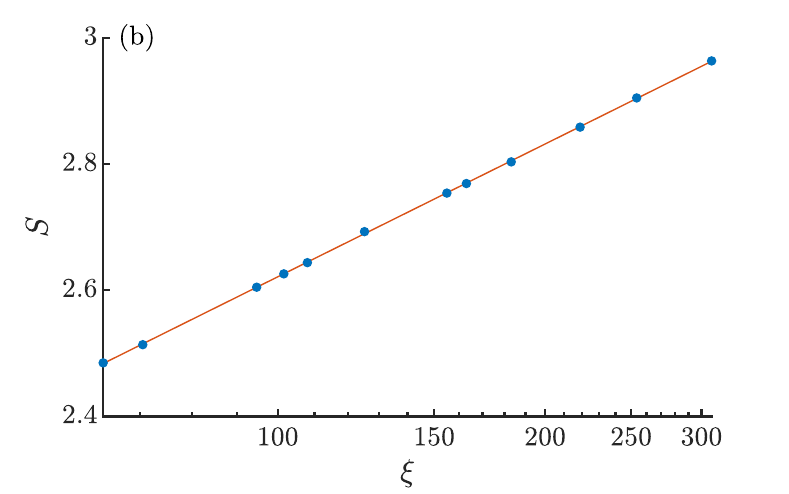}}
	\caption{Scaling of the entanglement entropy as a function of correlation length in the $\RP$ model for temperatures (a) $T = 0.4$ and (b) $T = 0.3$. The dashed line in (a) represents the extrapolated correlation length at this temperature. The line in (b) corresponds to a fit of the scaling form Eq.~\eqref{eq:cardy}.}
	\label{fig:RP2scaling}
\end{figure}

\begin{figure}
	\subfigure{\includegraphics[width=\columnwidth]{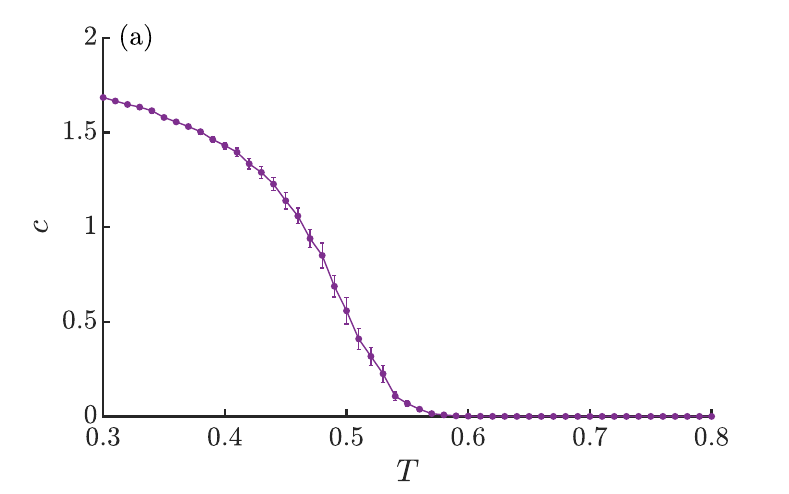}}
	\subfigure{\includegraphics[width=\columnwidth]{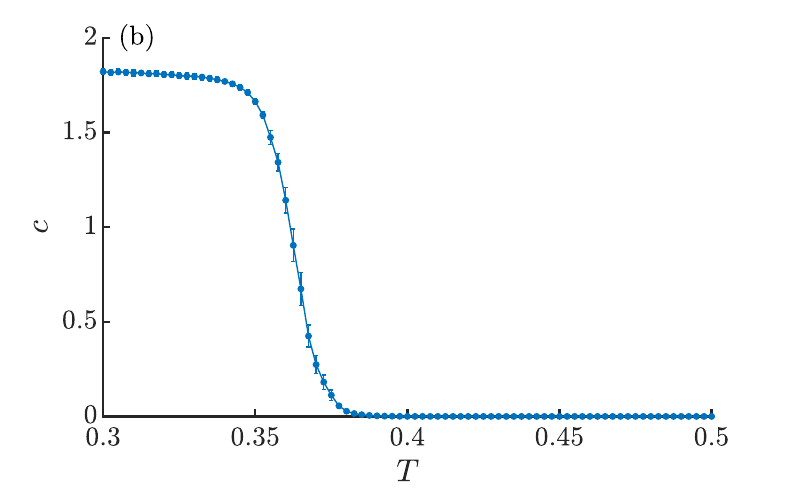}}
	\caption{Effective central charge as a function of temperature in the (a) Heisenberg model and the (b) $\RP$ model.}
	\label{fig:c}
\end{figure}

We begin by considering the entanglement-entropy scaling in the $\RP$ model at temperatures $T = 0.4$ and $T = 0.3$, chosen respectively above and below the sudden increase in correlation length diagnosed in the previous section. The results are depicted in Fig.~\ref{fig:RP2scaling}. At the higher temperature $ T = 0.4 $ the entanglement entropy quickly saturates with increasing bond dimension. This is consistent with fact that the correlation length at this temperature is fairly small, indicating a gapped transfer matrix. At the lower temperature $ T = 0.3 $ we observe a strong agreement with the scaling form \eqref{eq:cardy}, yielding an effective central charge $ c = 1.82 \pm 0.01 $. To determine whether this scaling indicates true criticality or rather corresponds to a pseudo-critical region, we study the change in scaling behavior with changing temperature. Indeed, for a system exhibiting a phase transition from a gapped high-temperature phase to a critical low-temperature phase, such as the XY model, one would observe an abrupt onset of scaling behavior separating a high-temperature region with $c = 0$ from a low-temperature region with a constant $c$ corresponding to the CFT describing the critical phase \cite{vanderstraeten2019approaching}. Once again, we contrast the scaling behavior as a function of temperature for the $\RP$ model with that of the Heisenberg model. Fig.~\ref{fig:c} shows the effective central charge obtained from \eqref{eq:cardy} as a function of the temperature for both models. An effective central charge $ c = 0 $ indicates a saturating entanglement entropy, whereas a finite value is consistent with the occurrence of scaling behavior. This characterization again reveals a fundamental difference between the two models.

For the Heisenberg model we observe a gradual onset of scaling behavior around $ T = 0.55 $, after which the effective central charge increases steadily as the temperature is lowered further. This picture does clearly not correspond to that of a critical low-temperature phase as sketched above, consistent with our results in the previous section. We argue that the observed pseudo-critical scaling is caused by the proximity to the zero-temperature $\Othree$ fixed point. Indeed, the onset of scaling behavior in the top panel of Fig.~\ref{fig:c} is very sensitive to the bond dimensions considered when extrapolating the effective central charge using \eqref{eq:cardy}. Namely, we observe that when increasing the maximal bond dimension taken into account when fitting the effective central charge, the onset of scaling behavior is shifted towards lower temperatures. We therefore conjecture that if one would use results for ever increasing bond dimensions, the effective scaling curve for the Heisenberg model would keep shifting towards lower temperatures until only the true $T = 0$ critical point remains, with central charge $c = 2$ on account of the two Nambu-Goldstone modes associated with the spontaneous symmetry breaking which occurs at zero temperature \cite{calabrese2004entanglement}. This conclusion is supported by studies of the bilinear-biquadratic Heisenberg chain which report a similar phenomenon when approaching the $\SU(3)$ point \cite{hu2014berryphaseinduced, dai2022absence}. 

Just as with the correlation length, the $ \RP $ model exhibits a much more abrupt behavior. At the onset of scaling behavior the effective central charge increases sharply towards a seemingly stable plateau around $ c \approx 1.8 $ as the temperature decreases further. As stated above, this behavior could be consistent with a transition towards a critical low-temperature phase. However, just as with the correlation-length divergence there are some obstructions towards such a conclusion. Firstly, the effective central charge does not assume a single fixed value in the low-temperature region, but in fact increases very slightly as the temperature is lowered further. In addition, the specific value $ c \approx 1.8 $ is not expected from any known field theoretic low energy description of the $ \RP $ model. Indeed, from perturbation theory one would only expect a zero-temperature fixed point with $c = 2$. As such, just as before we conclude that the observed behavior is not consistent with true criticality, but rather the signature of a pseudo-critical region. In App.~\ref{sec:cutoff} we again provide evidence that the observed scaling is not tainted by cutoff or finite bond dimension effects.

It is important to note that, even though the observed scalings in the Heisenberg and $\RP$ models have been diagnosed as arising from pseudo-criticality, there is a fundamental difference between the two models. Indeed, while for the Heisenberg model we have observed that an increase in maximal bond dimension leads to a shift of the effective scaling curve towards lower temperatures, this is not the case for the $\RP$ model. We found that the effective scaling curve for the $\RP$ model is robust against an increase in the maximal bond dimension (cf. also App.~\ref{sec:cutoff}). This observation is consistent with the definite crossover temperature found in the previous section, and could again be explained by the proximity to a nearby fixed point outside the model parameter space.

\begin{figure}
\includegraphics[width=\columnwidth]{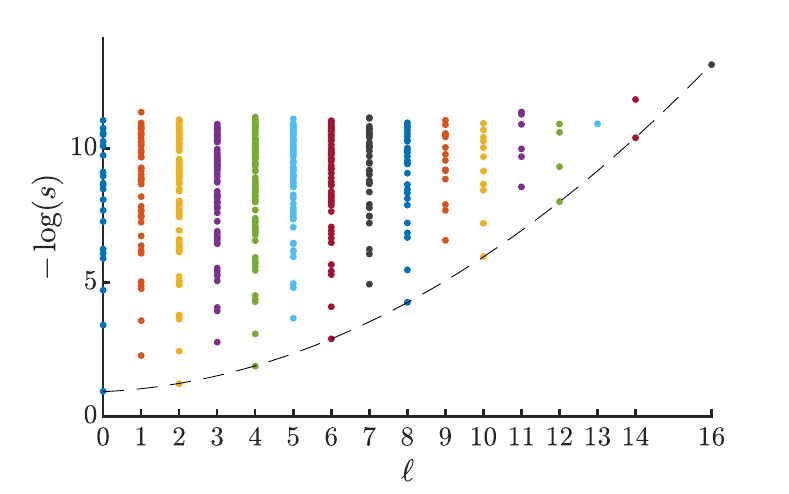}
\caption{Entanglement spectrum of a fixed point MPS with $ D = 4780 $ for the $ \RP $ transfer matrix at $T = 0.3$. The dashed line shows a quadratic envelope of the form $a \ell(\ell+1) + b$.}
\label{fig:spectrum}
\end{figure}

\begin{figure}
\subfigure{\includegraphics[width=\columnwidth]{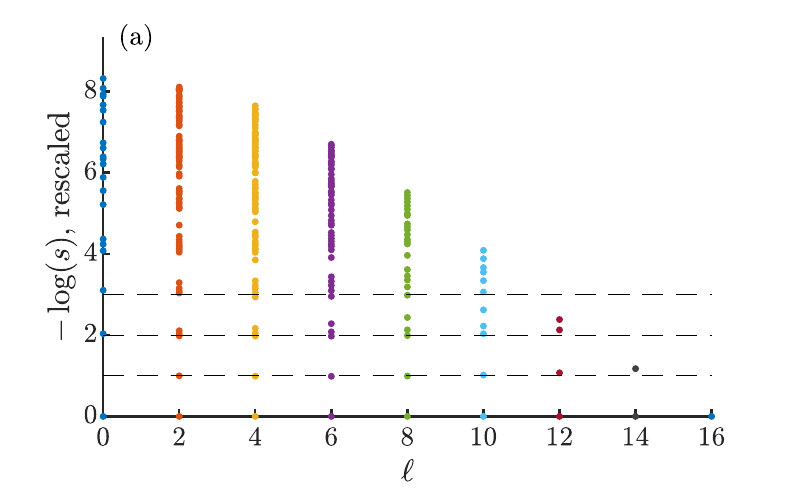}}
\subfigure{\includegraphics[width=\columnwidth]{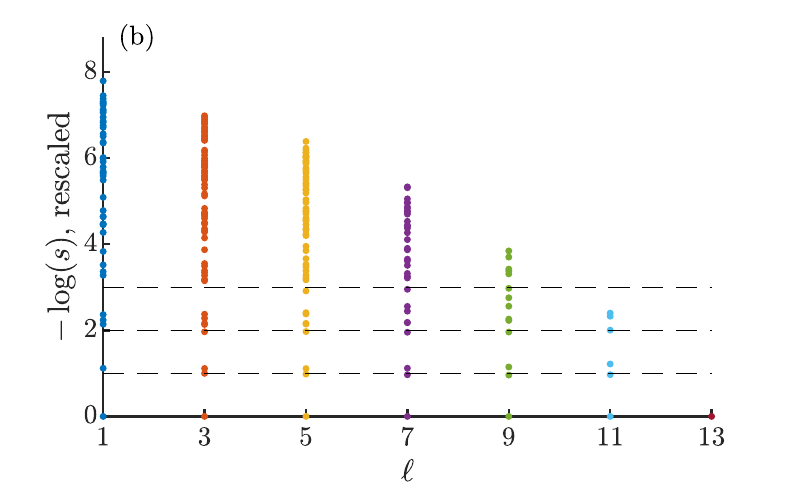}}
\caption{Entanglement spectrum from Fig.~\ref{fig:spectrum} split into (a) even and (b) odd integer spin sectors. After a shift of each sector followed by a global rescaling, we obtain equidistant spectra for even and odd integer spin sectors respectively. The dashed lines indicate approximate energy levels.}
\label{fig:spectrumSplit}
\end{figure}

As a final characterization of the low-temperature region in the $ \RP $ model we investigate the entanglement spectrum of the MPS fixed point. As pointed out in Ref.~\onlinecite{lauchli2013operator}, in a true critical phase the low-lying part of the entanglement spectrum of a bipartition of the MPS fixed point should correspond to the energy spectrum of a boundary CFT. In Ref.~\onlinecite{vanderstraeten2019approaching} such a boundary-CFT spectrum was found with high precision in the MPS fixed point of the critical XY model. In Fig.~\ref{fig:spectrum} we show the entanglement spectrum of a fixed-point MPS with bond dimension $ D = 4780 $ for the $ \RP $ model at $ T = 0.3 $. As we impose $ \SOthree$ symmetry on the MPS, the spectrum is labeled by the angular momenta appearing at the virtual level. We immediately note that the lowest-lying branch follows a quadratic envelope of the form $a \ell(\ell+1) + b$, which is indeed a hallmark of a boundary-CFT spectrum. For a true boundary-CFT spectrum, however, a shift of the different sectors followed by a rescaling with an overall energy gap should yield an equidistant spectrum where each level exhibits a specific degeneracy. This procedure is unsuccessful for the spectrum in Fig.~\ref{fig:spectrum}: the levels corresponding to even and odd integer spins do not coincide, and they exhibit different degeneracies. By splitting up the even and odd integer spin sectors we do however obtain equidistant spectra, as depicted in Fig.~\ref{fig:spectrumSplit}.

In fact, the obtained low-temperature entanglement spectrum bears a strong resemblance to spectra encountered in quantum spin systems with nematic order \cite{penc2011spin, hu2014berryphaseinduced}, which are inherently characterized by an alternating pattern of even and odd-integer spins. We may therefore conclude that the low-temperature region of the $\RP$ model carries a strong signature of nematic QLRO. The slight deviations in the spectrum from what one would expect from true nematic QLRO further reinforce our assessment that the low-temperature region in fact exhibits pseudo-critical behavior at length scales below the true correlation length. This conclusion is supported by the findings of Ref.~\cite{hu2014berryphaseinduced}, where similar signatures of nematic QLRO at length scales below the true correlation length were reported due to the proximity to the $\SU(3)$ point.

\section{Discussion and outlook}
\label{sec:discussion}

In this paper we have made use of state-of-the-art tensor-network methods to perform a comparative study of the two-dimensional classical Heisenberg and $ \RP $ models, with the main goal of providing a fresh viewpoint on the question whether the $\RP$ model exhibits a finite-temperature phase transition to a quasi-long-range ordered low-temperature phase. In particular, using uniform MPS with explicit $\SOthree$ symmetry directly in the thermodynamic limit, we were able to probe (i) correlation lengths up to $\cO(10^3)$ sites accurately, (ii) the scaling of entanglement entropy, and (iii) universal features of MPS entanglement spectra, the latter of which are inaccessible in finite-size Monte Carlo approaches.

For the Heisenberg model we have found no signs of a finite-temperature phase transition, supporting the scenario of asymptotic freedom. In the low-temperature region, we have observed an effective scaling of the entanglement entropy on length scales that are small compared to the true correlation length in the system. A much more abrupt onset of scaling behavior was observed in the $\RP$ model, hinting towards a finite-temperature phase transition where the divergence of the correlation can be fitted to the characteristic BKT form. A more careful analysis has shown, however, that the divergence of the correlation length softens for values around a few hundred sites. The scaling of the entanglement entropy and the MPS entanglement spectra were shown to exhibit strong signatures of nematic quasi-long-range order, but were again found to be inconsistent with true criticality. Therefore, our findings are in agreement with the scenario of a pseudo-critical region \cite{catterall1998nature, paredesv2008no, farinas-sanchez2010critical, tomita2014finitesize} for the $\RP$ model. While the scaling of the entanglement entropy in both models was diagnosed as a signature of pseudo-criticality, our results reveal a fundamental difference in scaling behavior between the Heisenberg and $\RP$ models. While the onset of scaling in the Heisenberg model depends on the bond dimensions considered, this onset occurs at a temperature that is bond-dimension independent in the $\RP$ model, indicating that this crossover temperature is an inherent property of the model.

Given the similarity of the issues encountered in studies of the two-dimensional $\RP$ model and the fully frustrated Heisenberg antiferromagnet on the triangular lattice, it could prove worthwhile to also investigate the latter model with tensor-network methods. While the framework used in this work was detailed for the specific case of the square lattice, it can be readily generalized to lattices of arbitrary coordination number. When combined with recently developed methods for tackling frustration in statistical mechanics models using tensor networks \cite{vanhecke2021solving}, this may provide new insights into the physics of the triangular-lattice antiferromagnet as well.

In the future, it will be interesting to consider more general interaction terms within the tensor-network representation of $\Othree$ models. Indeed, the transformation in Eq.~\ref{eq:expansion} from the $\Othree$ group basis to the basis of irreducible representations is an example of a duality transformation \cite{lootens2021categorytheoretic}. In that context, it would be interesting to search for models exhibiting discrete holomorphicity \cite{fendley2021integrability}, which might yield integrable models within the $\Othree$ universality class. Additionally, we can search for additional, possibly complex, couplings that would drive the pseudo-critical region towards a true critical point.

The success of the tensor-network approach for diagnosing criticality in classical models with continuous symmetries serves as a motivation for investigating more exotic types of criticality. For example, we could study the critical behavior of the surface of a three-dimensional model. Here, we can first capture the bulk using projected entangled-pair states \cite{vanderstraeten2018residual}, after which we can use the methods detailed in this work to simulate the two-dimensional surface physics. Recent works on surface critical behavior for $\ON$ models suggest that a variety of new exotic scaling phenomena can be expected \cite{metlitski2022boundary, padayasi2022extraordinary, parisentoldin2022boundary}.

During the preparation of this work we learned of a related study of the two-dimensional classical Heisenberg and $\RP$ models by Ueda and Oshikawa \cite{ueda2022tensor} using tensor-network renormalization. Their conclusions largely coincide with ours where the studies overlap.


\begin{acknowledgments}
We acknowledge inspiring discussions with Jutho Haegeman, Atsushi Ueda, Masaki Oshikawa, Philipp Schmoll, Matteo Rizzi, Andreas L\"auchli and Rui-Zhen Huang, and are grateful to Atsushi Ueda and Masaki Oshikawa for sharing their draft \cite{ueda2022tensor} with us prior to publication. This work was supported by the Research Foundation Flanders (FWO) via grants 11H7223N and FWO20/PDS/115, and by the European Research Council (ERC) under the European Unions Horizon 2020 research and innovation programme (Grant No. 647905 (QUTE)).
\end{acknowledgments}


\bibliography{bibliography}

\appendix


\section{Framework}
\label{sec:framework}

We consider a system of three-dimensional classical spins $ \vec{s}_i $ of unit length placed on the sites of a two-dimensional square lattice $ \cL $, which interact according to the Hamiltonian
\begin{equation}\label{eq:ham}
    \cH = -\sum_{\braket{ij}} \left (\vec{s}_i \cdot \vec{s}_j\right )^p\,.
\end{equation}
Here $ p = 1 $ and $ p = 2 $ correspond to the Heisenberg and $ \RP $ models respectively, and $ \braket{ij} $ labels all links of the lattice.

\begin{widetext}

\subsection{Partition function}
\label{sec:partitionFunction}

The partition function for the system \eqref{eq:ham} is given by
\begin{equation}
	\cZ = \left (\prod_{i} \int \frac{\txd\Omega_i}{4\pi}\right ) \left (\prod_{\braket{ij}} \txe^{\beta\left(\vec{s}_i \cdot \vec{s}_j\right)^p}\right )\,,
\end{equation}
where $ \beta = 1/T $ represents the inverse temperature and
\begin{align}
	\int \txd \Omega_i &= \int_{0}^{\pi} \sin \theta_i \txd \theta_i \int_{0}^{2\pi} \txd \phi_i\,, \\
	\vec{s}_i \cdot \vec{s}_j &= \sin \theta_i \sin \theta_j \left ( \cos \phi_i \cos \phi_j + \sin \phi_i \sin \phi_j \right ) + \cos \theta_i \cos \theta_j\,.
\end{align}
In order to write this partition function as a tensor network, we perform a character expansion of the Boltzmann weights in terms of spherical harmonics \cite{savit1980duality, liu2013exact}
\begin{equation}\label{eq:expExpansion}
	\txe^{\beta\left(\vec{s}_i \cdot \vec{s}_j\right)^p} = \sum_{\ell} f_{\ell}(\beta) \sum_{m = -\ell}^{\ell} \bar{Y}_{\ell m}(\theta_i, \phi_i) Y_{\ell m}(\theta_j, \phi_j)\,,
\end{equation}
where the expansion coefficients $ f_\ell(\beta) $ are defined in terms of Legendre polynomials $ P_\ell(x) $,
\begin{equation}\label{eq:flbis}
	f_\ell(\beta) = 2\pi \int_{-1}^{1} \txd x P_\ell(x) \txe^{\beta x^p} \,.
\end{equation}
This expansion gives rise to an expression for the partition function of the form
\begin{equation}\label{eq:Zdiscretized}
	\cZ = \sum_{\{\ell_i\}, \{m_i\}} \left ( \prod_{i \in \cL} f_{\ell_i}(\beta) \right )  \left ( \prod_{s \in \cL} F_{\ell_1 m_1, \ell_2 m_2}^{\ell_3 m_3, \ell_4 m_4} \right ) \,,
\end{equation}
where the products run over all links $ i $ and sites $ s $ of the lattice.
The factors $ F_{\ell_1 m_1, \ell_2 m_2}^{\ell_3 m_3, \ell_4 m_4} $ for every site are given by
\begin{equation}\label{eq:Fint}
	F_{\ell_1 m_1, \ell_2 m_2}^{\ell_3 m_3, \ell_4 m_4} = \int \frac{\txd\Omega_i}{4\pi} Y_{\ell_1 m_1}(\theta, \phi) Y_{\ell_2 m_2}(\theta, \phi) \bar{Y}_{\ell_3 m_3}(\theta, \phi) \bar{Y}_{\ell_4 m_4}(\theta, \phi).
\end{equation}
Using the expression for the fusion of two spherical harmonics in terms of Wigner $ 3j $ symbols
\begin{equation}\label{eq:Yprod}
	Y_{\ell_1 m_1}(\theta, \phi) Y_{\ell_2 m_2}(\theta, \phi) = \sqrt{\frac{(2\ell_1+1)(2\ell_2+1)}{4\pi}}
	\sum_{k, n} (-1)^n \sqrt{2k + 1}
	\threej{\ell_1}{\ell_2}{k}{0}{0}{0}
	\threej{\ell_1}{\ell_2}{k}{m_1}{m_2}{-n}
	Y_{k n}(\theta, \phi)
\end{equation}
combined with their orthogonality we obtain
\begin{equation}\label{eq:F}
	F_{\ell_1 m_1, \ell_2 m_2}^{\ell_3 m_3, \ell_4 m_4} = \frac{1}{4\pi}\sum_{k, n}
	G(\ell_1, \ell_2, k, m_1, m_2, n)
	G(\ell_3, \ell_4, k, m_3, m_4, n)\,.
\end{equation}
Here we have defined a modified Gaunt coefficient $ G $ associated with each fusion or splitting vertex of angular momenta,
\begin{align}\label{eq:G}
	G(\ell_1, \ell_2, k, m_1, m_2, n) 
	&= (-1)^n
	\sqrt{\frac{(2\ell_1+1)(2\ell_2+1)(2k+1)}{4\pi}}
	\threej{\ell_1}{\ell_2}{k}{0}{0}{0}
	\threej{\ell_1}{\ell_2}{k}{m_1}{m_2}{-n} \nonumber \\
	&= (-1)^{\ell_1 - \ell_2}
	\sqrt{\frac{(2\ell_1+1)(2\ell_2+1)}{4\pi}}
	\threej{\ell_1}{\ell_2}{k}{0}{0}{0}
	\braket{\ell_1 m_1, \ell_2 m_2| k n} \,,
\end{align}
where we have used the relation between Wigner $ 3j $ symbols and Clebsch-Gordan coefficients in the last line.

By introducing a local four-leg tensor $ O $,
\begin{equation}\label{eq:Odef}
    \diagram{q1}{2} = 
    \left( \prod_{i=1}^{4} f_{\ell_i}(\beta) \right)^{\frac{1}{2}} F_{\ell_1 m_1, \ell_2 m_2}^{\ell_3 m_3, \ell_4 m_4} \,,
\end{equation}
we can write the partition function \eqref{eq:Zdiscretized} as the contraction of a tensor network,
\begin{equation}
	\cZ = \quad \diagram{q1}{1} \;.
\end{equation}
Note that for the $ \RP $ model this tensor only contains even integer spins on its legs, as $ f_\ell(\beta) = 0 $ for any odd integer $ \ell $ in this model.

\end{widetext}

\subsection{Transfer matrix}

The corresponding row-to-row transfer matrix
\begin{equation}\label{eq:Tbis}
	T(\beta) = \quad \diagram{q1}{3}
\end{equation}
can be viewed as an operator acting on an infinite one-dimensional chain. The value of the partition function is therefore entirely determined by the leading eigenvalue $ \Lambda(\beta) $ of this operator which scales as the number of sites per row, $ \Lambda(\beta) = \lambda(\beta)^{N_x} $. The corresponding eigenvector $ \ket{\Psi} $ is referred to as the fixed point of the transfer matrix,
\begin{equation}\label{eq:Teig}
	T(\beta)\ket{\Psi} = \Lambda(\beta) \ket{\Psi}\,.
\end{equation}
As the transfer matrix \eqref{eq:Tbis} is Hermitian by definition of the tensor \eqref{eq:Odef}, the eigenvalue problem can be reformulated in terms of the optimization of a corresponding free energy density,
\begin{equation}\label{eq:psi}
	\ket{\Psi} = \argmin_{\ket{\Psi}} \left( -\frac{1}{\beta} \frac{1}{N_x} \frac{\braket{\Psi|T(\beta)|\Psi}}{\braket{\Psi|\Psi}} \right)\,.
\end{equation}
For the infinite system at hand, we approximate the fixed point of the transfer matrix by a uniform MPS characterized by a single tensor $ A $,
\begin{equation}
\ket{\Psi(A)} = \quad \diagram{q1}{4} \,.
\end{equation}
The optimal $ A $ can then be characterized in terms of the variational problem
\begin{equation}
	\max_A \frac{\braket{\Psi(\bar{A})|T(\beta)|\Psi(A)}}{\braket{\Psi(\bar{A})|\Psi(A)}}\,,
\end{equation}
which we can solve efficiently using the VUMPS algorithm \cite{zauner-stauber2018variational, fishman2018faster, vanderstraeten2019tangentspace}. The corresponding free energy density can be computed as
\begin{equation}\label{eq:f}
	f(\beta) = -\frac{1}{\beta} \log \lambda(\beta)\,,
\end{equation}
where $ \lambda $ is the leading eigenvalue of the channel operator
\begin{equation}\label{eq:lambda}
	\lambda = \rho_{\text{max}} \left ( \diagram{q1}{5} \right)
\end{equation}
and we assume a normalized MPS
\begin{equation}\label{eq:norm}
	\rho_{\text{max}} \left ( \diagram{q1}{6} \right) = 1\,.
\end{equation}

\subsection{Symmetries}

The models \eqref{eq:ham} possess a global $ \Othree $ symmetry: they are invariant under any global rotation or reflection of the spins. In the tensor-network representation, this global symmetry is reflected by the fact that the tensor \eqref{eq:Odef} is invariant under the transformation
\begin{equation}\label{eq:RO}
	\diagram{q1}{7} = \diagram{q1}{8}\,,
\end{equation}
where $ U_g $ is the representation of an arbitrary rotation or reflection $ g \in \Othree $. Specifically, an arbitrary rotation $ g = \cR $ is represented by the matrix
\begin{equation}\label{eq:R}
	\diagram{q1}{9} = \delta_{\ell_1}^{\ell_2} D^{\ell_1}_{m_1 m_2}(\cR)\,,
\end{equation}
with $ D^{\ell} $ the irreducible representation of $ \SOthree $ corresponding to spin $ \ell $. Similarly, a reflection $ g = \cP $ is represented by
\begin{equation}\label{eq:P}
	\diagram{q1}{10} = \delta_{\ell_1}^{\ell_2} \delta_{m_1}^{m_2} (-1)^{\ell_1}\,.
\end{equation}
While the global rotation symmetry follows automatically from the Clebsch-Gordan coefficients contained within \eqref{eq:F}, the additional global reflection symmetry is enforced by the symmetry properties of the $ 3j $ symbols appearing in \eqref{eq:G}. The tensor-network representation of the partition function therefore explicitly exhibits the full $ \Othree $ symmetry comprised of rotations and reflections. We note that for the $ \RP $ model the additional local reflection symmetry is also manifestly preserved in this representation. Since the local tensor \eqref{eq:Odef} only contains even integer spins in this model, it is trivially invariant under the action of \eqref{eq:P} on any of its legs.

As the breaking of a global continuous symmetry is prohibited by the Mermin-Wagner theorem, the fixed point MPS of the transfer matrix must also be invariant under any transformation of the form,
\begin{align}\label{eq:RA}
	\diagram{q1}{11} &= \nonumber \\
	&\diagram{q1}{4}\,.
\end{align}
The fundamental theorem of MPS \cite{perez-garcia2007matrix} then allows to associate this $ \SOthree $ invariance to a symmetry property of the local MPS tensor
\begin{equation}\label{eq:RAv}
	\diagram{q1}{12} = \diagram{q1}{13}\,,
\end{equation}
where $ v_g $ is a (possibly projective) representation of $ \SO(3) $. Specifically, this means that the MPS must transform according to a representation of $ \SU(2) $ on the virtual level. Imposing the symmetry property \eqref{eq:RAv} enforces the MPS tensor to have a certain block structure where each block is labeled by the irreducible representations of $ \SU(2) $ on each leg of the MPS tensor. As the tensor \eqref{eq:Odef} also has this structure, all of the numerics can be carried out within the framework of $ \SU(2) $ symmetric tensors \cite{mcculloch2002nonabelian, singh2010tensor, weichselbaum2012nonabelian}. In this framework a tensor is explicitly stored in a block diagonal form, where each block is split up into a degeneracy part and a concatenation of Clebsch-Gordan coefficients which corresponds to the fusion tree associated to that block. For the local tensor \eqref{eq:Odef} for example, this degeneracy tensor is readily obtained by dividing out the two Clebsch-Gordan coefficients that occur in the expression \eqref{eq:F}. By automatically accounting for symmetry constraints through efficient manipulations of fusion trees, this block diagonal structure can be exploited to greatly improve the efficiency of numerical simulations, allowing access to effective bond dimensions that are out of reach for conventional dense tensor-network approaches.

As a final remark, we note that the legs of the local tensor \eqref{eq:Odef} in principle have an infinite dimension. For the purpose of numerical simulations, these indices must be truncated at a certain cutoff value $ \lmax $ for the charge on all legs. This does not result in a significant loss of accuracy, as the coefficients \eqref{eq:flbis} decay rapidly when increasing $ \ell $ for sufficiently large temperatures. As stated in Sec.~\ref{sec:method}, all results in the main text were obtained using $ \lmax = 5 $ for the Heisenberg model and $ \lmax = 6 $ for the $ \RP $ model. At very low temperatures however, the decay of the $ f_\ell(\beta) $ occurs more slowly, and therefore more charges should be taken into account to accurately represent the true untruncated model. A motivation for the validity of our angular momentum cutoff for the temperature range considered in the main text is given in App.~\ref{sec:cutoff}. While very-low-temperature features cannot be captured in a quantitatively accurate manner for a strict cutoff, it turns out that the actual qualitative features are unchanged in heavily truncated versions of the model. This is also demonstrated in App.~\ref{sec:cutoff}.

\subsection{Local observables and correlation functions}
\label{sec:obs}

Consider now a single-site observable $ g(\vec{s}_\mu) $ which only depends on the spin at site $ \mu $. Its expectation value would take the form
\begin{equation}\label{eq:exp1}
	\braket{g} = \frac{1}{\cZ} \left (\prod_{i} \int \frac{\txd\Omega_i}{4\pi}\right ) \left ( g(\vec{s}_\mu) \txe^{-\beta E(\{\vec{s}_i\})}\right ).
\end{equation}
Such a quantity can be readily computed by placing a modified local tensor at site $ \mu $ by incorporating the factor $ g(\vec{s}_\mu) $ in the integration \eqref{eq:F} over the angles at that site. However, for any nontrivial single-site observable $ g(\vec{s}_\mu) $ will not be invariant under rotations. As such, the corresponding modified tensor cannot be constructed using the symmetric framework. In addition, the expectation value of such a tensor will always be identically zero when evaluated using a manifestly symmetric boundary MPS. Similarly, any correlation function that is not invariant under global $ \SOthree $ transformations will identically vanish within our framework, consistent with the Mermin-Wagner theorem.

The interesting observables are therefore those which obey global rotation invariance. An important class of such observables are two-point correlation functions of the form $ h(\vec{s}_\mu \cdot \vec{s}_\nu) $ which depend only on the inner product of spins located at sites $ \mu $ and $ \nu $. Their expectation value takes the form
\begin{equation}\label{eq:exp2}
	\braket{h} = \frac{1}{\cZ} \left (\prod_{i} \int \frac{\txd\Omega_i}{4\pi}\right ) \left ( h(\vec{s}_\mu \cdot \vec{s}_\nu) \txe^{-\beta E(\{\vec{s}_i\})} \right ).
\end{equation}
By performing a character expansion of $ h(\vec{s}_\mu \cdot \vec{s}_\nu) $ in exactly the same way as \eqref{eq:flbis} and adding the additional spherical harmonics to the integration over angles at sites $ \mu $ and $ \nu $, this expectation value can be computed by means of two modified local tensors
\begin{widetext}
\begin{subequations}\label{eq:Hs}
\begin{align}\label{eq:H+}
	\diagram{q1}{14} \quad = \left( \prod_{i \neq 3} f_{\ell_i}(\beta) \right)^{\frac{1}{2}} \sqrt{h_{\ell_3}} \left (M^+\right )_{\ell_1 m_1, \ell_2 m_2, \ell_3 m_3}^{\ell_4 m_4, \ell_5 m_5} \;, \\
	\label{eq:H-}
	\diagram{q1}{15} \quad = \left( \prod_{i \neq 3} f_{\ell_i}(\beta) \right)^{\frac{1}{2}} \sqrt{h_{\ell_3}} \left (M^-\right )_{\ell_1 m_1, \ell_2 m_2}^{\ell_3 m_3, \ell_4 m_4, \ell_5 m_5} \;,
\end{align}
\end{subequations}
where
\begin{equation}\label{eq:hl}
	h_\ell = 2\pi \int_{-1}^{1} \txd x P_\ell(x) h(x) \,,
\end{equation}
and
\begin{subequations}\label{eq:Ms}
\begin{align}\label{eq:M+}
	\left (M^+\right )_{\ell_1 m_1, \ell_2 m_2, \ell_3 m_3}^{\ell_4 m_4, \ell_5 m_5}
	=
	\frac{1}{4\pi} \sum_{k_1, n_1} \sum_{k_2, n_2} G(\ell_1, \ell_2, k_1, m_1, m_2, n_1) \, G(k_1, \ell_3, k_2, n_1, m_3, n_2) G(\ell_4, \ell_5, k_2, m_4, m_5, n_2)\;, \\
	\label{eq:M-}
	\left (M^-\right )_{\ell_1 m_1, \ell_2 m_2}^{\ell_3 m_3, \ell_4 m_4, \ell_5 m_5}
	=
	\frac{1}{4\pi} \sum_{k_1, n_1} \sum_{k_2, n_2} G(\ell_1, \ell_2, k_1, m_1, m_2, n_1) \, G(\ell_3, k_2, k_1, m_3, n_2, n_1) G(\ell_4, \ell_5, k_2, m_4, m_5, n_2) \;,
\end{align}
\end{subequations}
\end{widetext}
Using the same boundary MPS that was used to contract the partition function, any such two-point correlation function can then be evaluated as
\begin{equation}
	\braket{h} = \frac{\left (\diagram{q1}{16}\right )}{\left (\diagram{q1}{17}\right )}.
\end{equation}
For the models \eqref{eq:ham} the most relevant correlation functions are the spin-spin correlation functions,
\begin{equation}\label{eq:scor}
	h(\vec{s}_\mu \cdot \vec{s}_\nu) = (\vec{s}_\mu \cdot \vec{s}_\nu)^p,
\end{equation}
which, given \eqref{eq:hl}, are captured in terms of tensors \eqref{eq:Hs} where the additional leg carries an angular momentum $ \ell = 1 $ and $ \ell = 0, 2 $ for the Heisenberg and $ \RP $ models respectively. In addition, we note that for the $ \RP $ model any correlation function $ h(\vec{s}_\mu \cdot \vec{s}_\nu) $ that contains only odd integer powers of the spin inner product must vanish. Indeed, any odd integer power in the correlation function would, through the character expansion, lead to odd integer values of the spin $ \ell_3 $ on the additional leg of the tensors \eqref{eq:Hs}. However, all other legs carry only even integer spin in the $ \RP $ model. Thus, by the symmetry properties of the $ 3j $ symbols appearing in the modified Gaunt coefficients \eqref{eq:G} the tensors \eqref{eq:Hs} must be identically zero in such a case. This means that any two-point correlation function that does not respect the local reflection symmetry of the $ \RP $ model vanishes by construction.

For the specific case of two-point correlators between neighboring sites, the character expansion of the correlation function can be combined with that of the Boltzmann weight across the same link. This allows to evaluate the corresponding expectation value using a single tensor. For example, for the energy per link $ E(\vec{s}_\mu \cdot \vec{s}_\nu) = -(\vec{s}_\mu \cdot \vec{s}_\nu)^p $ associated to a horizontal link connecting neighboring sites $ \mu $ and $ \nu $ this procedure leads to a tensor of the form
\begin{equation}\label{eq:E}
	\diagram{q1}{18} = 
	\left( \prod_{i=1}^{3} f_{\ell_i}(\beta) \right)^{\frac{1}{2}} \frac{e_{l_4}(\beta)}{\sqrt{f_{l_4}(\beta)}} F_{\ell_1 m_1, \ell_2 m_2}^{\ell_3 m_3, \ell_4 m_4} \,,
\end{equation}
where
\begin{equation}\label{eq:el}
	e_l(\beta) = -2\pi \int_{-1}^{1} \txd x P_l(x) \, x^p \, \txe^{\beta x^p} \,.
\end{equation}
The energy per link can therefore be evaluated as
\begin{equation}\label{eq:energyEvaluation}
	\braket{E} = \frac{\left( \diagram{q1}{19} \right)}{\left( \diagram{q1}{20} \right)}.
\end{equation}

As a final remark, we note that one can easily obtain conventional order parameters from the tensor-network representation of two-point correlation functions. For the specific example of spin-spin correlation function, we define the local expectation value $ M_{\ell m} $ as
\begin{equation}\label{eq:Mlm}
	\diagram{q1}{21} = \quad \frac{\left( \diagram{q1}{22} \right)}{\left( \diagram{q1}{20} \right)}\,.
\end{equation}
From this object we can extract an order parameter $ M $ as
\begin{equation}\label{eq:M}
	M = \sqrt{\sum_{m=-p}^{p} \left (M_{pm}\right )^2}\,,
\end{equation}
which is a measure for the amount of long-range order present in the system. For the Heisenberg model ($ p = 1 $) \eqref{eq:Mlm} is equivalent to the conventional magnetization, while for the $ \RP $ model ($ p = 2 $) it corresponds to the nematic order parameter \cite{lebwohl1972nematicliquidcrystal}. By the Mermin-Wagner theorem these order parameters must vanish everywhere, but they are used in some Monte Carlo studies where finite system sizes allow for some degree of symmetry breaking. Similarly, while these order parameters are always identically zero when evaluated within the symmetric tensor framework, they can assume a nonzero value if symmetries are not explicitly imposed.


\section{Representations on the virtual MPS level}
\label{sec:spt}

\begin{figure}
\subfigure{\includegraphics[width=\columnwidth]{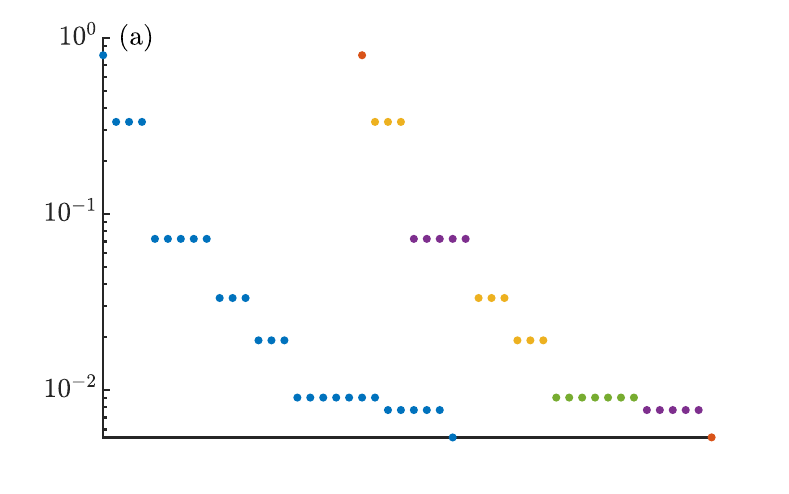}}
\subfigure{\includegraphics[width=\columnwidth]{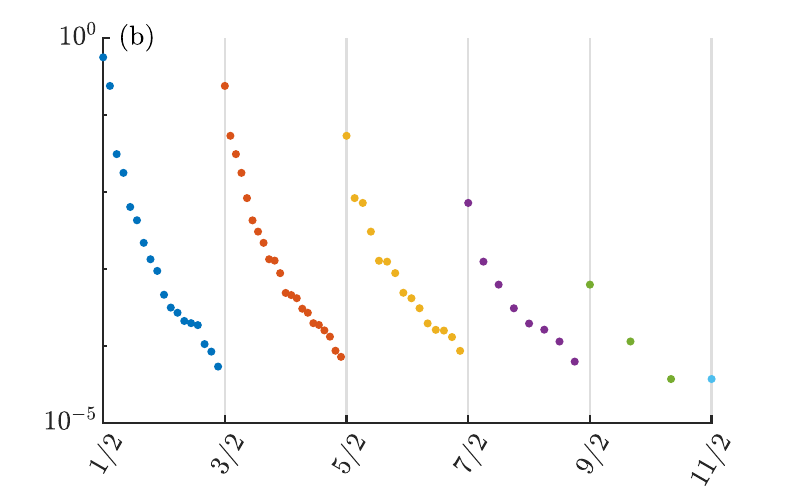}}
\caption{Entanglement spectra of fixed points for the Heisenberg model at $T = 0.6$. (a) Spectra of fixed point MPSs (left) without and (right) with imposing $\SOthree$ symmetry. The different integer spin sectors in the symmetric spectrum are plotted in different colors. (b) Spectrum for a fixed point MPS with half-integer spins on the virtual level.}
\label{fig:spt}
\end{figure}

Here we motivate our choice to exclusively use integer spin charges at the virtual MPS level in the simulations presented in the main text. While in principle both integer and half-integer spin charges can occur on the virtual level of the MPS used to contract the partition function $\eqref{eq:Zdiscretized}$, it is fairly simple to exclude the half-integer case in the current context. As a first diagnostic, in Fig.~\ref{fig:spt}(a) we depict the entanglement spectrum of two MPSs of moderate bond dimension $D = 28$ which were optimized for the Heisenberg model at $T = 0.6$. For one MPS we impose no symmetries while for the other we impose $\SOthree$ symmetry using only integer spins on the virtual level. Whereas the nonsymmetric spectrum can in principle take any form, the symmetric spectrum will by construction display odd degeneracies corresponding to integer spin charges. The exact coincidence of the two spectra in Fig.~\ref{fig:spt}(a) clearly proves that the true fixed point contains only integer spins on the virtual level. Fig.~\ref{fig:spt}(b) shows the spectrum of a fixed point MPS optimized at the same temperature using half-integer spins on the virtual level. We immediately see that the entire spectrum apart from the largest value is degenerate. This can be understood from the fact that the corresponding MPS attempts to imitate an MPS with only integer spins on the virtual level by consistently matching pairs of half-integer spins in the spectrum. Exactly the same behavior as that shown in Fig.~\ref{fig:spt} is observed for the $\RP$ model.

For low temperatures the characterization is less straightforward, as a uniform MPS of finite bond dimension on which no symmetries are imposed will spontaneously break a global continuous symmetry when close to a critical point. As the correlation lengths at lower temperatures are enormous in both models under consideration here, a similar phenomenon takes place. As such, we can no longer compare to a nonsymmetric fixed point in order to determine the virtual structure of the true fixed point. However, it turns out that we can still exclude half-integer spins by simple arguments. First, an MPS with half-integer spins on the virtual level systematically requires a larger bond dimension to achieve the same truncation error compared to an MPS with integer spins on the virtual level, implying that it is an inferior variational ansatz. Second, when using half-integer charges on the virtual level the fixed points systematically converge to a noninjective MPS for all temperatures. This implies that these fixed points are unphysical for the models considered in this work.


\section{Universality and the effect of the cutoff}
\label{sec:cutoff}

\begin{figure}
\includegraphics[width=\columnwidth]{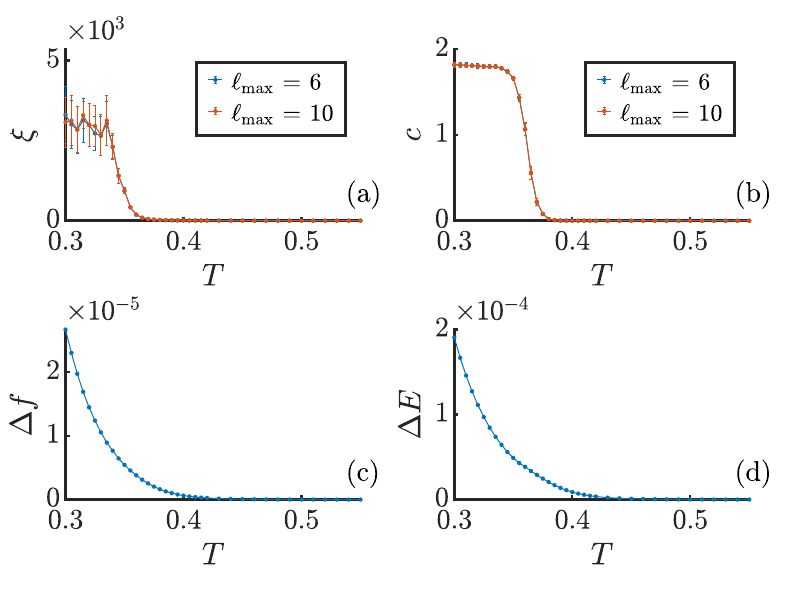}
\caption{Comparison of results for the $\RP$ model with $\lmax = 6$ and $\lmax = 10$. (a,b) The extrapolated correlation length and effective central charge as a function of temperature. (c, d) The relative difference in the free energy density and the energy per link between both cutoffs as a function of temperature.}
\label{fig:RP2compare}
\end{figure}

\begin{figure}
\includegraphics[width=\columnwidth]{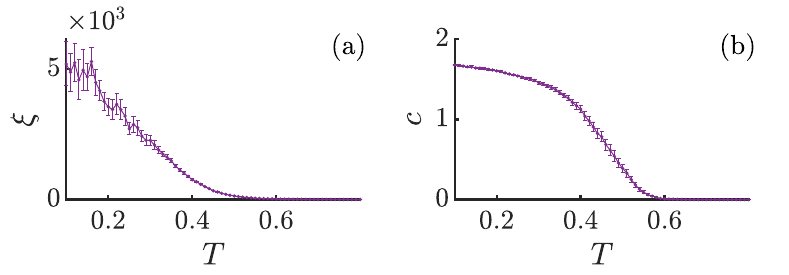}
\caption{Results for the Heisenberg model with $\lmax = 2$ and a maximal MPS bond dimension $D = 3000$, showing (a) the extrapolated correlation length and (b) the effective central charge as a function of temperature.}
\label{fig:heis_N2}
\end{figure}

\begin{figure}
\includegraphics[width=\columnwidth]{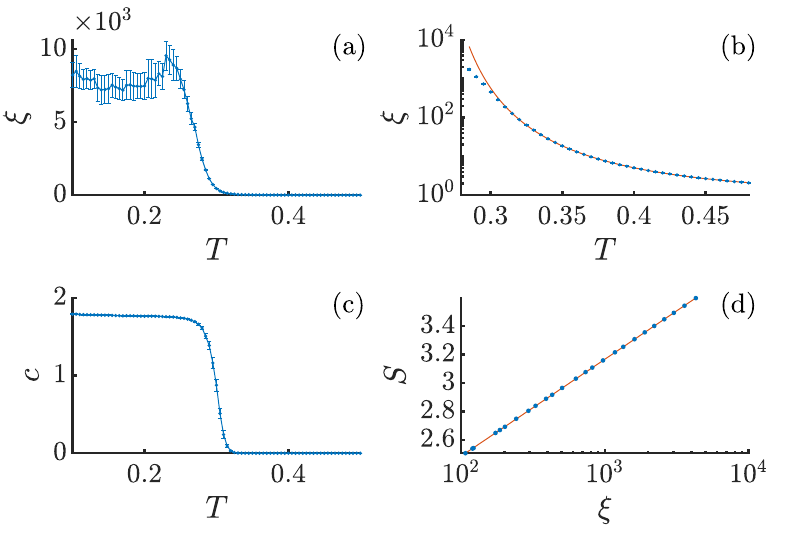}
\caption{Results for the $\RP$ model with $\lmax = 2$ and a maximal MPS bond dimension $D = 3000$. (a) The extrapolated correlation length as a function of temperature. (b) A fit of the BKT scaling form \eqref{eq:BKTfit} to the correlation length divergence. (c) The effective central charge as a function of the temperature. (d) The entanglement-entropy scaling at $T = 0.15$.}
\label{fig:RP2_N1}
\end{figure}

In this Appendix we provide evidence that the conclusions drawn in the main text are not tainted by the effect of the approximations made. First, we motivate that the angular momentum cutoff $\lmax = 5$ and $\lmax = 6$ used in the main text for the Heisenberg and $\RP$ models respectively are sufficient to capture the true behavior of the untruncated model. To this end, we repeat numerics for the $\RP$ model for a selection of temperatures using a larger cutoff $\lmax = 10$, and compare the results to the $\lmax = 6$ case, as shown in Fig.~\ref{fig:RP2compare}. The values of local observables such as the free energy density and the energy per link display a larger relative difference as the temperature is lowered, up to values $\sim10^{-4}$. This is to be expected from the fact that the expansion coefficients \eqref{eq:flbis} decay more slowly with increasing angular momentum as the temperature is lowered. This is also the reason why $T = 0.3$ was systematically chosen as the lower bound for the temperature ranges considered in the main text, as the error on local observables due to truncation would quickly become too large below this temperature. Despite minor differences in local quantities, it can be seen from the top panels of Fig.~\ref{fig:RP2compare} that the extrapolated quantities for both values of the cutoff exactly coincide. This proves that the results in the main text accurately characterize the universal scaling behavior of the true model. A similar diagnostic for the Heisenberg model leads to exactly the same conclusion.

Next we show that the nature of the observed scaling behavior is not affected by the truncation error due to finite bond dimensions. To this end we repeat numerics for more severely truncated versions of both the Heisenberg and $\RP$ models using a cutoff $\lmax = 2$, but with an enlarged maximal bond dimension of $D = 3000$. The results are shown in Fig.~\ref{fig:heis_N2} and Fig.~\ref{fig:RP2_N1} for the Heisenberg and $\RP$ models respectively. We immediately note that the qualitative behavior of both the correlation length and the effective central charge looks identical to the results obtained with larger cutoff values, but shifted towards lower temperatures. This implies two things. First, it means that the universal behavior is not changed by the cutoff in the representation of the partition function. This was to be expected, as the truncated models possess the exact same symmetry as the true untruncated models. Second, it shows that our conclusions were not affected by the finite MPS truncation error, as we observe the same behavior using a larger maximal bond dimension. In particular, we emphasize the right panels of Fig.~\ref{fig:RP2_N1}. The top-right panel shows the exact same behavior as Fig.~\ref{fig:RP2kt}, namely an initial approach to a true BKT divergence followed by the softening of the exponential increase. The bottom-right panel shows the scaling of the entanglement entropy with correlation length at $T = 0.15$ for a maximal MPS bond dimension of $D = 2 \cdot 10^{4}$, exhibiting perfect agreement with the scaling form \eqref{eq:cardy} up to correlation lengths $\xi > 4 \cdot 10^3$. This all provides strong evidence that our conclusions are unaffected by the approximations made.

\end{document}